# Capturing dynamics and thermodynamics of a three-level quantum heat engine via programmable quantum circuits

Gao-xiang Deng[a], Zhe He[b], Yu Liu[a], Wei Shao[b*], Zheng Cui[b,**]

a Institute of Thermal Science and Technology, Shandong University, Jinan, 250061, P.R. China

b Shandong Institute of Advanced Technology, Jinan, 250100, P.R. China

*Correspondence: shao@sdu.edu.cn;

**Correspondence: zhengc@sdu.edu.cn

**Abstract**

This research employs the Kraus representation and Sz.-Nagy dilation theorem to model a three-level quantum heat on quantum circuits, investigating its dynamic evolution and thermodynamic performance. The feasibility of the dynamic model is validated by tracking the changes of population. On the basis of reinforcement learning algorithm, the optimal cycle of the quantum heat engine for maximal average power is proposed and verified by the thermodynamic model. The stability of quantum circuit simulations is scrutinized through a comparative analysis of theoretical and simulated results, predicated on an orthogonal test. These results affirm the practicality of simulating quantum heat engines on quantum circuits, offering potential for substantially curtailing the experimental expenses associated with the construction of such engines.

# 1 Introduction

The concept of Quantum Heat Engine (QHE) was first proposed in the 1950s and 1960s [1, 2]. However, due to the limitations of theoretical framework at that time, the QHE research did not garner significant attentions. Then, the maturation of open quantum system theory [3, 4] and quantum thermodynamics [5] led to a surge in the QHE research over the past two decades [6].

A prevailing question in the QHE researches is whether the quantum effects can bolster the performance. Early theoretical research suggested that the manifestation of quantum effects, such as quantum coherence and entanglement, would consume resources, thereby diminishing the performance-a phenomenon termed "quantum friction" [7-11]. Conversely, other theoretical studies have indicated that quantum effects, including quantum coherence [12-17], quantum entanglement [18-20], energy level degeneracy [17, 21, 22], quantum fluctuations [23-27], and multi-body collaboration [28-31], could potentially enhance the QHE performance. These theoretical discrepancies have spurred scholars to conduct experimental investigations.

Thanks to the advancements in low-temperature and laser technology, researchers have successfully constructed various types of QHEs in laboratories. These QHEs utilize a range of elements, including cold atoms [32, 33], ions [34-37], electrons [38-41], photons [42, 43], particle pair spins [44-47], and negatively charged nitrogen vacancy ($NV^-$) centers in diamond [48-50], among others. Many experiments have demonstrated that quantum effects can indeed enhance the performance of heat engines [37, 40, 47, 48, 50]. For instance, Ref. [47] found that when the thermal stroke time is shorter than the system decoherence time, the output work of the QHE increases. However, these experiments are often costly and time-consuming.

In the other side, the advent of cloud quantum technology, such as those offered by IBM and Microsoft, provides a cost-effective solution for simulating real quantum models. This approach involves constructing a quantum circuit model locally, submitting it to the cloud for execution, and awaiting the returned results. This method significantly reduces experimental costs while enabling the study of actual quantum model performance. Currently, many quantum models [51-61], including QHEs [51, 54, 57, 58, 61], have been successfully implemented as quantum circuits and have yielded promising results on quantum computers.

This study chooses the three-level QHE as it is the smallest model capable of "autonomous operation" [62-65]. The benefit of autonomous operation is the allowance of the heat engine cycle to operate independently and elimination of reliance on external resources. For instance, the efficiency improvement attributed to the squeeze bath [66, 67] was discovered to have a hidden cost, as it consumes external non-thermal resources [68]. Thus this article is structured as follows. Initially, the physical model of the three-level QHE, the corresponding quantum circuit models, and reviews of related prior works are introduced. Subsequently, using the population evolution and average power as indicators evaluates the quantum circuit models. Then, employing an orthogonal test, the stability of the quantum circuit simulation results is verified by steady-state power and efficiency. Lastly, the study concludes with a summary of the findings.

## 2 Models and Methods

This section unfolds as follows: Initially, the physical model and parameters of the QHE are delineated followed by the introduction of the quantum circuit models, which include dynamics model for population simulation, thermodynamic model for average power evaluation, and error mitigation method. Then the reinforcement learning algorithm based optimal cycle and the orthogonal test based steady-state performance analysis of the QHE are provided. The section ends with a summary of the methods and tools utilized in this study.

### 2.1 Physical Model

Fig. 1 gives the transitions between energy levels and corresponding thermodynamic processes of a three-level QHE.

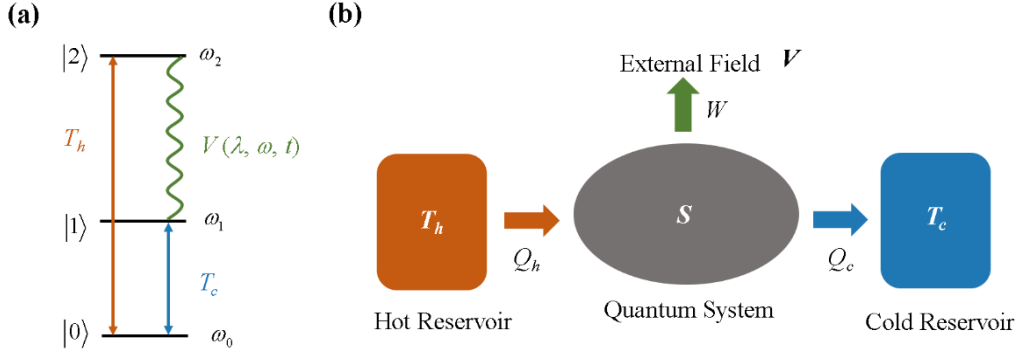

Fig. 1. Schematic of a three-level quantum heat engine (QHE). (a) Energy levels and transitions. The eigenstates of the quantum system's free Hamiltonian are denoted as $|0\rangle$, $|1\rangle$, and $|2\rangle$ while their corresponding eigenfrequencies are $\omega_0$, $\omega_1$, and $\omega_2$. $|0\rangle$ and $|2\rangle$ are coupled to the hot reservoir at $T_h$, and $|0\rangle$ and $|1\rangle$ are coupled to the cold reservoir at $T_c$. The transitions between $|1\rangle$ and $|2\rangle$ are induced by the external field $V(t) = \lambda e^{i\omega t}|1\rangle\langle 2| + \lambda e^{-i\omega t}|2\rangle\langle 1|$. $\lambda$, $\omega$, and $t$ are the intensity, frequency, and evolution time, respectively. (b) Thermodynamic processes. The interaction with the hot reservoir, cold reservoir, and external field results in heat absorption $Q_h$, heat release $Q_c$, and work output $W$, respectively.

The dynamics of the quantum system $S$ within this QHE is governed by the Gorini-Kossakowski-Lindblad-Sudarshan (GKLS) equation [69],

$$\partial_t \rho_S = \mathcal{L}\rho_S = -i[H_S, \rho_S] + \mathcal{D}(\rho_S(t)) \quad (1)$$

with the time-varying Hamiltonian of the quantum system $H_S$

$$H_S = \begin{pmatrix} \omega_0 & 0 & 0 \\ 0 & \omega_1 & \lambda e^{i\omega t} \\ 0 & \lambda e^{-i\omega t} & \omega_2 \end{pmatrix} \quad (2)$$

Here, $\mathcal{L}$ is the generator and $\mathcal{D}$ is the dissipator, symbolizing the heat dissipation with the reservoirs. Detailed information regarding the governing equation, i.e., Eq. (1), and heat-work relations can be found in Refs. [70, 71]. The parameters of the QHE utilized in the subsequent simulation are presented in Table 1.

Table 1. Parameters of the reservoirs and external field for the three-level QHE [70]. Here, $\beta = 1/k_B T$ is the inversed temperature of the reservoir, and $k_B$ is Boltzmann's constant. $g_1$ and $g_2$ are coupling functions, which respectively represent the coupling to the cold and hot reservoir, $\varepsilon_{21} = \varepsilon_2 - \varepsilon_1$.

| Parameters | $\beta_c \omega_{10}$ | $\beta_h \omega_{10}$ | $\omega_{20}/\omega_{10}$ | $\lambda/\omega_{10}$ | $\omega/\omega_{10}$ |
|---|---|---|---|---|---|
| Value | 5 | 1 | 2.5 | 0.5 | $\dfrac{\varepsilon_{21}^2 + \dfrac{1}{4}(g_1 + g_2)^2}{\omega_2 - \omega_1}$ |

## 2.2 Quantum Circuit Models

Simulating QHEs on quantum computers through quantum circuits requires only a simple compilation which reduces the construction costs as well as enabling the investigation of practical performance and potential applications.

### 2.2.1 Dynamics Simulation

Dynamic model generally describes the transition state of the physical system, i.e. predicts the final state from its initial state of the system by governing equations. In the context of QHEs, the investigation of dynamics encompasses the representation of the quantized system state and modeling of evolution governed by the master equation [69]. Consequently, to simulate the dynamics of a QHE on a quantum circuit, two issues must be addressed: the state representation and evolution modeling.

Firstly, **state representation**. The system state of a QHE can be represented by a density matrix as,

$$\rho = \sum_{i=0}^{n-1} p_i |\varphi_i\rangle\langle\varphi_i| \qquad (3)$$

where, $\rho$, $n$, $|\varphi_i\rangle$, and $p_i$ denote the system state, system size, pure state, and probability, respectively. This decomposition can be achieved through the eigen-decomposition of the system state. Additionally, the supplementary material provided in Ref. [72] offers a method to flatten the density matrix. This technique enables direct simulation of the system state without decomposition, but at the cost of increased dimension.

To facilitate the resolution of the GKLS equation, a diagonalization of the Hamiltonian in Eq. (2) is performed,

$$H_S = \sum_n \varepsilon_n |\varepsilon_n\rangle\langle\varepsilon_n| \tag{4}$$

where the eigenvalues $\varepsilon_n$ are given by,

$$\varepsilon_0 = \omega_0$$
$$\varepsilon_1 = \frac{\omega_1 + \omega_2}{2} - \sqrt{\left(\frac{\omega_2 - \omega_1}{2}\right)^2 + \lambda^2} \tag{5}$$
$$\varepsilon_2 = \frac{\omega_1 + \omega_2}{2} + \sqrt{\left(\frac{\omega_2 - \omega_1}{2}\right)^2 + \lambda^2}$$

and corresponding eigenstates $|\varepsilon_n\rangle$ are,

$$|\varepsilon_0\rangle = |0\rangle$$
$$|\varepsilon_1\rangle = |1\rangle\cos\frac{\theta}{2} - |2\rangle e^{-i\omega t}\sin\frac{\theta}{2} \tag{6}$$
$$|\varepsilon_2\rangle = |1\rangle e^{i\omega t}\sin\frac{\theta}{2} + |2\rangle\cos\frac{\theta}{2}$$

Simultaneously, the expected values of the system density matrix's diagonal elements are defined as follows:

$$\rho_{00} = \langle\varepsilon_0|\rho_S|\varepsilon_0\rangle$$
$$\rho_{11} = \langle\varepsilon_1|\rho_S|\varepsilon_1\rangle \tag{7}$$
$$\rho_{22} = \langle\varepsilon_2|\rho_S|\varepsilon_2\rangle$$

Secondly, **evolution modeling**. Given the initial state, the final state of a QHE can typically be determined by the GKLS equation. The corresponding evolution can be represented by the generator $\mathcal{L}$, as illustrated in Eq. (1). In the quantum circuits, quantum gates are added to simulate the evolution from the initial state to the final state. However, the quantum circuits require the added gates unitary operators. Operators that do not satisfy unitarity, such as $\mathcal{L}$, cannot be directly incorporated into quantum circuits as gates. Therefore, to simulate the evolution of the QHE on the quantum circuits, the following steps need to be undertaken:

(1) Kraus Representation. This involves decomposing the evolution of the QHE into the sum of Kraus operators:

$$\rho(t) = \sum_k \mathbf{M}_k \rho(0) \mathbf{M}_k^\dagger \tag{8}$$

Here, $\rho(0)$ represents the given initial density matrix, $\rho(t)$ is the final density matrix, and $\mathbf{M}_k$ is referred to as the Kraus operator, which satisfies the following relationship:

$$\sum_k \mathbf{M}_k^\dagger \mathbf{M}_k = \mathbf{I} \tag{9}$$

. $\rho(0)$ and $\rho(t)$ can be diagonalized as

$$\begin{aligned}\Lambda(0) &= \mathbf{U}(0)^\dagger \rho(0) \mathbf{U}(0) \\ \Lambda(t) &= \mathbf{U}(t)^\dagger \rho(t) \mathbf{U}(t)\end{aligned} \tag{10}$$

Here, $\Lambda(0)$ and $\Lambda(t)$ are the diagonal matrices of the initial and final density matrices, respectively, $\mathbf{U}(0)$ and $\mathbf{U}(t)$ are the corresponding unitary transformation matrices. Substituting these into Eq. (8) gives the Kraus representation of the final density matrix as follows:

$$\begin{aligned}\rho(t) &= \sum_k \mathbf{M}_k \rho(0) \mathbf{M}_k^\dagger \\ &= \sum_k \left(\mathbf{U}(t)\mathbf{M}_k'\mathbf{U}(0)^\dagger\right) \rho(0) \left(\mathbf{U}(t)\mathbf{M}_k'\mathbf{U}(0)^\dagger\right)^\dagger\end{aligned} \tag{11}$$

Here,

$$\mathbf{M}_k = \mathbf{U}(t)\mathbf{M}_k'\mathbf{U}(0)^\dagger \tag{12}$$

And $\mathbf{M}_k'$ can be written based on the final density matrix, its specific form can be found in Ref. [73]. Additionally, it's also feasible to recast the GKLS equation using the Kraus representation [74].

(2) Sz.-Nagy Dilation. According to the Sz.-Nagy dilation theorem, any contraction operator $\mathbf{A}$ can be dilated into the following unitary operator [75]:

$$\mathbf{U_A} = \begin{pmatrix} \mathbf{A} & \mathbf{D}_{\mathbf{A}^\dagger} \\ \mathbf{D_A} & -\mathbf{A}^\dagger \end{pmatrix} \tag{13}$$

where $\mathbf{D_A} = \sqrt{\mathbf{I} - \mathbf{A}^\dagger \mathbf{A}}$ is the defect operator of $\mathbf{A}$. A contraction operator is defined as an operator, denoted by $\mathbf{A}$, that satisfies the following relationship for any non-zero vector $\mathbf{v}$:

$$\|\mathbf{A}\| = \sup \frac{\|\mathbf{A} \cdot \mathbf{v}\|}{\|\mathbf{v}\|} \leq 1 \tag{14}$$

By leveraging the contraction property of Kraus operators [72], the Kraus operators in Eq. (8) and (12) can be dilated into unitary operators, as outlined in Eq. (13).

(3) Quantum Circuit Simulation. Through Eq. (13), a contraction operator $\mathbf{A}$ of dimension $n \times n$ can be converted into a unitary operator $\mathbf{U}_\mathbf{A}$ of dimension $2n \times 2n$. However, it necessitates modification,

$$\mathbf{U}_\mathbf{A}^{qc} = \begin{pmatrix} \mathbf{U}_\mathbf{A} & 0 & \cdots & 0 \\ 0 & 1 & \cdots & 0 \\ \vdots & \vdots & \ddots & \vdots \\ 0 & 0 & \cdots & 1 \end{pmatrix}_{2^N \times 2^N} \tag{15}$$

prior to its integration into the quantum circuit. Here, $N$ is the smallest integer such that satisfies $2^N \geq 2n$ and also the number of qubits of the quantum circuit.

In summary, to simulate the evolution from $\rho(0)$ to $\rho(t)$ on quantum circuits, the first step is to decompose $\rho(0)$ according to Eq. (3) and rewrite Eq. (8) as:

$$\rho(t) = \sum_k \mathbf{M}_k \rho(0) \mathbf{M}_k^\dagger = \sum_{i=0}^{n-1} p_i \sum_k \mathbf{M}_k |\varphi_i(0)\rangle\langle\varphi_i(0)| \mathbf{M}_k^\dagger \tag{16}$$

Then, based on Eqs. (12), (13), and (15), the Kraus operator $\mathbf{M}_k$, corresponding unitary operator $\mathbf{U}_{\mathbf{M}_k}$, and circuit unitary operator $\mathbf{U}_{\mathbf{M}_k}^{qc}$, will be defined, respectively. Finally, the simulation results of the quantum circuit can be expressed as,

$$\sum_{i=0}^{n-1} p_i |\varphi_i^{qc}(t)\rangle\langle\varphi_i^{qc}(t)| = \sum_{i=0}^{n-1} p_i \sum_k \mathbf{U}_{\mathbf{M}_k}^{qc} |\varphi_i^{qc}(0)\rangle\langle\varphi_i^{qc}(0)| \mathbf{U}_{\mathbf{M}_k}^{qc\dagger} \tag{17}$$

where, the initial and final pure states on the quantum circuit are given by,

$$\begin{aligned} |\varphi_i^{qc}(0)\rangle &= \left(|\varphi_i(0)\rangle^T, 0, \ldots, 0\right)_{2^N}^T \\ |\varphi_i^{qc}(t)\rangle &= \left(|\varphi_i(t)\rangle^T, 0, \ldots, 0\right)_{2^N}^T \end{aligned} \tag{18}$$

and corresponding density matrices are,

$$\rho^{qc}(0) = \sum_{i=0}^{n-1} p_i \left|\varphi_i^{qc}(0)\right\rangle\left\langle\varphi_i^{qc}(0)\right| = \begin{pmatrix} \rho(0) & 0 & \cdots & 0 \\ 0 & 0 & \cdots & 0 \\ \vdots & \vdots & \ddots & \vdots \\ 0 & 0 & \cdots & 0 \end{pmatrix}_{2^N \times 2^N}$$

$$\rho^{qc}(t) = \sum_{i=0}^{n-1} p_i \left|\varphi_i^{qc}(t)\right\rangle\left\langle\varphi_i^{qc}(t)\right| = \begin{pmatrix} \rho(t) & 0 & \cdots & 0 \\ 0 & 0 & \cdots & 0 \\ \vdots & \vdots & \ddots & \vdots \\ 0 & 0 & \cdots & 0 \end{pmatrix}_{2^N \times 2^N}$$

(19)

Simultaneously, to manage the relationship between the quantum circuit states and QHE system states, employing binary encoding and Eq. (18) to establish the following mapping:

$$\begin{aligned} |0\cdots 00\rangle_N &\to \left|\varphi_0^{qc}\right\rangle \to |\varphi_0\rangle \\ |0\cdots 01\rangle_N &\to \left|\varphi_1^{qc}\right\rangle \to |\varphi_1\rangle \\ &\vdots \\ |0\cdots\rangle_N &\to \left|\varphi_{n-1}^{qc}\right\rangle \to |\varphi_{n-1}\rangle \end{aligned}$$

(20)

For the QHE introduced in Section 2.1 ($n=3$), the corresponding mapping is given by,

$$\begin{aligned} |000\rangle &\to |\varepsilon_0\rangle \\ |001\rangle &\to |\varepsilon_1\rangle \\ |010\rangle &\to |\varepsilon_2\rangle \end{aligned}$$

(21)

Fig. 2 outlines the process of using quantum circuits to simulate the dynamics of a QHE. Specifically, Fig. 2(a) illustrates the simulation process for the system state, encompassing the decomposition of the initial state, transformation of evolution, and simulation of quantum circuits. Fig. 2(b) depicts a specific row of "Quantum circuit simulation" from Fig. 2(a), representing the pure state simulation process. The overall simulation results are presented in Fig. 2(c).

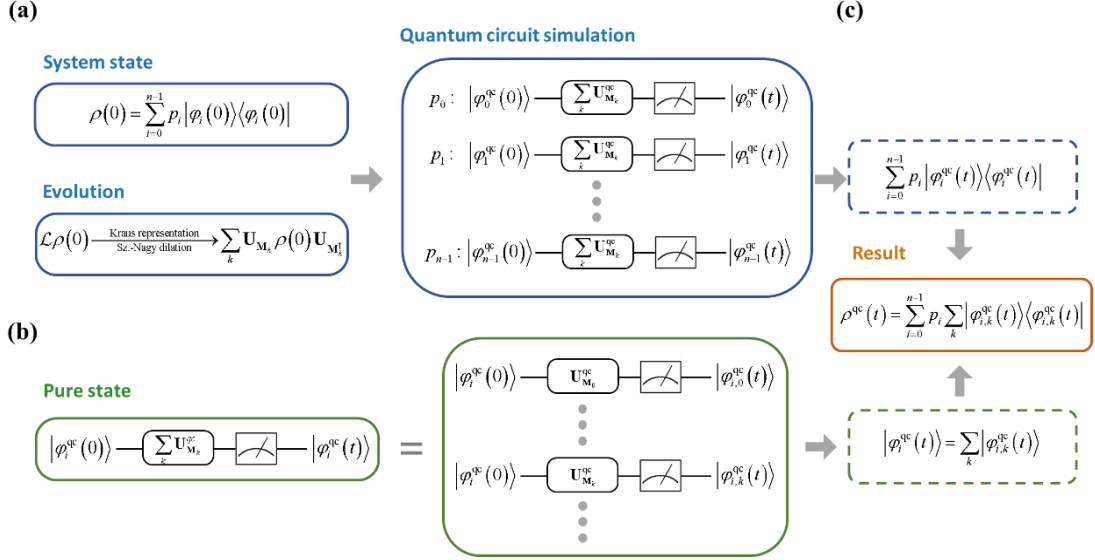

Fig. 2. Quantum circuit simulation of a QHE. (a) System state simulation. The upper-left blue box represents the decomposition of the initial state according to Eq. (3). Here, $\rho(0)$, $|\varphi_i(0)\rangle$, and $p_i$ are the initial system state, pure state and corresponding probability, respectively. The lower-left box shows the transformation of the generator $\mathcal{L}$, which denotes the evolution of QHE, into the unitary operators $\mathbf{U}_{\mathbf{M}_k}$ supported by the quantum circuit. The right-side blue box presents the quantum circuits model of the system state evolution. Each row encompasses the circuits model and corresponding simulation results depicted in panel (b). (b) Pure state simulation. For a pure state on quantum circuit $|\varphi_i^{qc}(0)\rangle$ in panel (a), the green box on the left represents the corresponding quantum circuits. The box on the right signifies the equivalent quantum circuit of each Kraus operator and corresponding simulation result $|\varphi_{i,k}^{qc}(t)\rangle$. (c) Simulation results. The upper blue and lower green dashed boxes denote the results of the system state simulation and pure state simulation, respectively. The comprehensive result is presented in the middle box, which is distinguished by an orange solid line. The variables depicted in the figure are elaborated from Eq. (16) to Eq. (20) in the main text.

### 2.2.2 Thermodynamics Evaluation

The performance analysis of QHEs involves determining thermodynamic quantities like internal energy, heat transfer, and work. This process requires the evaluation of the expected values of their corresponding observables. The evaluation of the expectation of an observable, $\langle \mathbf{O} \rangle = \mathrm{tr}(\mathbf{O}\rho)$, on quantum circuits can be divided into the following steps [72]:

(1) Reduction: This involves converting $\mathbf{O}$ into a contraction operator,

$$\tilde{\mathbf{O}} = \frac{\mathbf{O} + \mathbf{I}\|\mathbf{O}\|_{HS}}{2\|\mathbf{O}\|_{HS}} \qquad (22)$$

Here, $\|\mathbf{O}\|_{HS}$ denotes the Hilbert-Schmidt norm, which is the square root of the sum of the squares of the matrix elements.

(2) Cholesky decomposition:

$$\tilde{\mathbf{O}} = \mathbf{L}\mathbf{L}^\dagger \qquad (23)$$

where, $\mathbf{L}$ and $\mathbf{L}^\dagger$ are lower triangular and upper triangular matrices respectively, and the expectation of $\tilde{\mathbf{O}}$ can be expressed as,

$$\langle \tilde{\mathbf{O}} \rangle = \mathrm{tr}(\tilde{\mathbf{O}}\rho(t)) = \mathrm{tr}(\mathbf{L}\mathbf{L}^\dagger \rho(t)) = \mathrm{tr}(\mathbf{L}^\dagger \rho(t)\mathbf{L}) \qquad (24)$$

(3) Quantum circuit modeling: Using the Kraus representation from Eq. (8), Eq. (24) can be rewritten as,

$$\langle \tilde{\mathbf{O}} \rangle = \mathrm{tr}(\mathbf{L}^\dagger \rho(t)\mathbf{L}) = \mathrm{tr}\left(\mathbf{L}^\dagger \sum_k \mathbf{M}_k \rho(0) \mathbf{M}_k^\dagger \mathbf{L}\right) = \sum_k \mathrm{tr}(\mathbf{L}^\dagger \mathbf{M}_k \rho(0) \mathbf{M}_k^\dagger \mathbf{L}) \qquad (25)$$

Here, $\mathbf{M}_k$ is the Kraus operator. Consequently, the evaluation of $\langle \tilde{\mathbf{O}} \rangle$ can be executed by the evolution of $\rho(0)$, which is determined by the joint influence of operators $\mathbf{M}_k$ and $\mathbf{L}^\dagger$. To achieve this, the 2-dilation form of Sz.-Nagy dilation [75] needs to be applied,

$$\mathbf{U}_\mathbf{A} = \begin{pmatrix} \mathbf{A} & 0 & \mathbf{D}_{\mathbf{A}^\dagger} \\ \mathbf{D}_\mathbf{A} & 0 & -\mathbf{A}^\dagger \\ 0 & \mathbf{I} & 0 \end{pmatrix}, \mathbf{U}_\mathbf{B} = \begin{pmatrix} \mathbf{B} & 0 & \mathbf{D}_{\mathbf{B}^\dagger} \\ \mathbf{D}_\mathbf{B} & 0 & -\mathbf{B}^\dagger \\ 0 & \mathbf{I} & 0 \end{pmatrix} \qquad (26)$$

where, $\mathbf{A}$ and $\mathbf{B}$ are contraction operators. By taking $\mathbf{A} = \mathbf{M}_k$ and $\mathbf{B} = \mathbf{L}^\dagger$ into Eq. (26) and modifying the corresponding results in accordance with Eq. (15),

$$\begin{aligned}
\langle \tilde{\mathbf{O}} \rangle &= \sum_k \mathrm{tr}\left(\mathbf{L}^\dagger \mathbf{M}_k \rho(0) \mathbf{M}_k^\dagger \mathbf{L}\right) \\
&= \sum_k \mathrm{tr}\left(\mathbf{U}_{\mathbf{L}^\dagger}^{qc} \mathbf{U}_{\mathbf{M}_k}^{qc} \sum_i p_i |\varphi_i^{qc}(0)\rangle \langle \varphi_i^{qc}(0)| \mathbf{U}_{\mathbf{M}_k^\dagger}^{qc} \mathbf{U}_{\mathbf{L}}^{qc}\right) \\
&= \sum_i p_i \sum_k \mathrm{tr}\left(\mathbf{U}_{\mathbf{L}^\dagger}^{qc} \mathbf{U}_{\mathbf{M}_k}^{qc} |\varphi_i^{qc}(0)\rangle \langle \varphi_i^{qc}(0)| \mathbf{U}_{\mathbf{M}_k^\dagger}^{qc} \mathbf{U}_{\mathbf{L}}^{qc}\right)
\end{aligned} \quad (27)$$

(4) Expectation computation: According to Eq. (22),

$$\langle \mathbf{O} \rangle = \mathrm{tr}(\mathbf{O}\rho(t)) = \mathrm{tr}\left(\left(2\|\mathbf{O}\|_{HS}\tilde{\mathbf{O}} - \mathbf{I}\|\mathbf{O}\|_{HS}\right)\rho(t)\right) = 2\|\mathbf{O}\|_{HS}\langle \tilde{\mathbf{O}} \rangle - \|\mathbf{O}\|_{HS} \quad (28)$$

Substituting Eq. (27) into Eq. (28) will give the evaluation result of $\langle \mathbf{O} \rangle$ on quantum circuit.

### 2.2.3 Error Mitigation

In practical scenarios, the execution of a quantum circuit on a quantum computer is subject to various sources of error. These include quantum gate errors, environmental noise, time decoherence, and measurement errors, all of which can lead to discrepancies between the execution results and theoretical predictions. Thus the general error mitigation (GEM) [76] method is used to rectify these errors generated during the execution of quantum circuits. This technique calibrates the raw data by minimizing the following function:

$$f(x) = \sum_{i=1}^{2^N}\left(v_i - (\mathbf{G}\cdot\mathbf{x})_i\right)^2 \quad (29)$$

Here, $N$ represents the number of qubits, $v_i$ denotes the $i$ th element of raw data $\mathbf{v} = (v_1, v_2, \cdots, v_{2^N})^\mathrm{T}$, and $\mathbf{x} = (x_1, x_2, \cdots, x_{2^N})^\mathrm{T}$ is the calibrated data. $\mathbf{G}$ is the calibration matrix, the computation of which is detailed in Ref. [76].

### 2.3 Optimal cycle with maximal average power by reinforcement learning (RL) algorithm

In the previous work [77], a reinforcement learning (RL) algorithm was utilized to identify the optimal cycle for the three-level QHE. The objective is to maximize average power,

$$\langle P_{he} \rangle_i = (1-\gamma)\sum_{k=0}^{i}\gamma^k r_{i-k} \quad (30)$$

Here, $\gamma$ represents the discount factor, $i$ denotes the current training step, and the reward function,

$$r_{i+1} = \delta_{d,\bar{d}} \frac{1}{\Delta t} \Delta \langle E_S \rangle = \delta_{d,\bar{d}} \frac{1}{\Delta t} \left[ \text{tr}\left(\rho_S(t+\Delta t) H_S(t+\Delta t)\right) - \text{tr}\left(\rho_S(t) H_S(t)\right) \right] \quad (31)$$

where, $d = \{\text{hot, cold, work}\}$, $\bar{d} = \{\text{hot, cold}\}$, $t = i\Delta t$, and $\Delta t$ is the time step. The cycle is parameterized by a set,

$$a(t) = \{d(t), u(t)\} \quad (32)$$

where, discrete control parameter $d(t)$ determines the thermal process, and continuous control parameter $u(t)$ dictates the system's free Hamiltonian. A brief overview of this method is provided in Fig. 3 (a) (see Sec. II B.1 in Ref. [77] for more details) and the corresponding optimal cycle is given in Fig. 3 (b) (Fig. 5(c)).

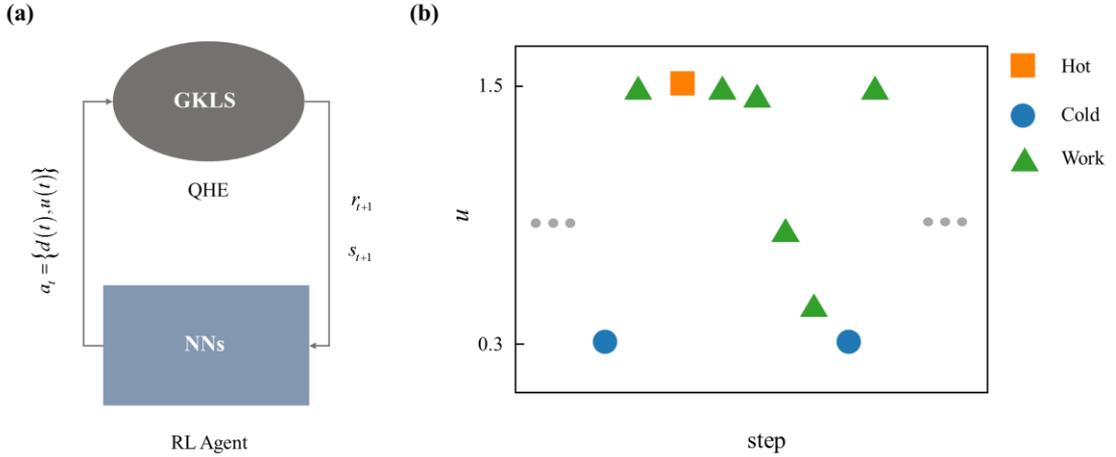

Fig. 3. (a) Reinforcement learning (RL) scheme for maximizing the average power of the three-level QHE. The RL agent, which is modeled by neural networks (NNs), controls the evolution of the QHE through $a(t) = \{d(t), u(t)\}$. The QHE, upon taking an action $a_t$ given by the RL agent, transitions to a new state $s_{t+1}$ according to the Gorini-Kossakowski-Lindblad-Sudarshan (GKLS) equation, as defined in Eq. (1). Simultaneously, it receives a reward $r_{t+1}$ as per Eq. (31). The NNs take $s_{t+1}$ and $r_{t+1}$ as input and subsequently output a new action. These steps are iteratively performed until convergence is achieved, which yields the optimal cycle with the maximum average power. (b) The optimal cycle of the QHE with maximal average power, as determined through RL. Specifically, it illustrates the internal relationship, denoted as

$a(t)$, between the number of cycle steps (time $t$), the thermal process $d(t)$, and the system state $u(t)$. In this representation, the abscissa corresponds to the number of cycle steps (time $t$), with different points and ordinates representing various thermal processes ($d$) and system states ($u$). Notably, the red square, blue circle, and green triangle symbolize the hot, cold, and work process, respectively.

The average power of the cycle (Fig. 5 (c)), obtained through RL, indeed outperforms other cycles (Fig. 5 (a), (b)) based on numerical calculations. However, this advantage is not clear when implemented on actual quantum devices. To this end, these cycles and their average power have been evaluated on quantum circuits, with the details provided in Section 2.2. Briefly, the expectation value of the Hamiltonian,

$$\langle H_S \rangle = \operatorname{tr}(\rho_S H_S) = \operatorname{tr}(H_S \rho_S) \tag{33}$$

in the context of the quantum circuit can be computed by substituting the operator **O** from Eq. (28) with $H_S$ from Eq. (2). Subsequently, the average power of different cycles running on quantum circuits are calculated using Eqs. (31) and (30). The corresponding results are presented in Section 3.2.

**2.4 Steady-state performance analysis by orthogonal test**

The steady-state performance of the three-level QHE is affected by several parameters [71], including temperatures $\beta_h$, $\beta_c$, coupling parameters $\gamma_h(\varepsilon_{20})$, $\gamma_c(\varepsilon_{10})$, $\gamma_h(\varepsilon_{10})$, $\gamma_c(\varepsilon_{20})$, external field intensity $\lambda$, and energy gap $\omega_{20}$. To efficiently and intuitively reveal the sensitivity of quantum circuit simulation results to these parameters, this section provides a brief introduction to the design of orthogonal tests and the computation of QHE's steady-state performance.

Existing research [71] has utilized orthogonal experiments to conduct an in-depth analysis of the theoretical steady-state characteristics of the three-level QHE. Therefore, this work will not delve into the details of the orthogonal test process and results. Instead, it focuses on assessing the reliability of quantum circuit simulations under different parameter configurations by comparing steady-state performance with existing theoretical results. To this end, power $P$ and efficiency $\eta$ are selected as indicators to evaluate the steady-state performance. The factors and corresponding levels are concurrently determined, as shown in Table 2. The L$_9$(3$^3$) orthogonal test, constructed based on Table 2, is presented in Table 3.

Table 2. Factors and levels used in the orthogonal test [71]. Here, $\Delta\beta$, $D_r$, and $D_d$ represent the temperature difference, resonant dissipation coupling parameters, and detuning dissipation coupling parameters, respectively.

| Level | Factor | | | | | |
|---|---|---|---|---|---|---|
| | $\Delta\beta$ | | $D_r$ | | $D_d$ | |
| | $\beta_h\omega_{10}$ | $\beta_c\omega_{10}$ | $\gamma_h(\varepsilon_{20})/\omega_{10}$ | $\gamma_c(\varepsilon_{10})/\omega_{10}$ | $\gamma_h(\varepsilon_{10})/\omega_{10}$ | $\gamma_c(\varepsilon_{20})/\omega_{10}$ |
| 1 | 1 | 5 | 0.5 | 0.5 | 0 | 0 |
| 2 | 0.5 | 2.5 | 1 | 1 | 0.5 | 0.5 |
| 3 | 0.2 | 1 | 2 | 2 | 2 | 2 |

Table 3. The order of L$_9$(3$^3$) orthogonal test. Cases 1-9 represent the order of the orthogonal test.

| Case | $\beta_h\omega_{10}$ | $\beta_c\omega_{10}$ | $\gamma_h(\varepsilon_{20})/\omega_{10}$ | $\gamma_c(\varepsilon_{10})/\omega_{10}$ | $\gamma_h(\varepsilon_{10})/\omega_{10}$ | $\gamma_c(\varepsilon_{20})/\omega_{10}$ |
|---|---|---|---|---|---|---|
| 1 | 1 | 5 | 0.5 | 0.5 | 0 | 0 |
| 2 | 1 | 5 | 1 | 1 | 2 | 2 |
| 3 | 1 | 5 | 2 | 2 | 0.5 | 0.5 |
| 4 | 0.5 | 2.5 | 0.5 | 0.5 | 2 | 2 |
| 5 | 0.5 | 2.5 | 1 | 1 | 0.5 | 0.5 |
| 6 | 0.5 | 2.5 | 2 | 2 | 0 | 0 |
| 7 | 0.2 | 1 | 0.5 | 0.5 | 0.5 | 0.5 |
| 8 | 0.2 | 1 | 1 | 1 | 0 | 0 |
| 9 | 0.2 | 2 | 2 | 2 | 2 | 2 |

In each case, $\lambda \in [0,1]$ and $\omega_{20} \in (1,5]$.

The steady-state power and efficiency are given by:

$$P^{SS} = J_h^{SS} + J_c^{SS} \quad (34)$$

$$\eta^{SS} = \frac{P^{SS}}{J_h^{SS}} \quad (35)$$

Here, the superscript "SS" denotes the "steady-state". And the steady-state heat flux (heat power) are:

$$J_h^{SS} = -\varepsilon_{10}\frac{1-\cos\theta}{2}\left[\gamma_h(\varepsilon_{10})\rho_{11}^{SS} - \gamma_h(-\varepsilon_{10})\rho_{00}^{SS}\right]$$
$$-\varepsilon_{20}\frac{1+\cos\theta}{2}\left[\gamma_h(\varepsilon_{20})\rho_{22}^{SS} - \gamma_h(-\varepsilon_{20})\rho_{00}^{SS}\right] \quad (36)$$

$$J_c^{SS} = -\varepsilon_{10}\frac{1+\cos\theta}{2}\left[\gamma_c(\varepsilon_{10})\rho_{11}^{SS} - \gamma_c(-\varepsilon_{10})\rho_{00}^{SS}\right]$$
$$-\varepsilon_{20}\frac{1-\cos\theta}{2}\left[\gamma_c(\varepsilon_{20})\rho_{22}^{SS} - \gamma_c(-\varepsilon_{20})\rho_{00}^{SS}\right] \quad (37)$$

where, $\rho_{00}^{SS}$, $\rho_{11}^{SS}$, and $\rho_{22}^{SS}$, correspond to the population of each energy level in the steady state. Notably, for the heat engine to function properly, the following relationship must be satisfied:

$$J_h^{SS} > 0,\ J_c^{SS} < 0,\ P^{SS} > 0 \quad (38)$$

For the specific interpretation of the parameters and the theoretical steady-state solution, please refer to Ref. [70].

**2.5 Methods and tools**

Given the initial state of the three-level QHE, the final density matrix is computed as per Eq. (1), and the corresponding quantum circuit model is established following the method outlined in Section 2.2. These quantum circuits are then simulated using *Qiskit* [78], a *Python* quantum computing framework provided by IBM. Specifically, choosing the qasm_simulator as the simulation backend and FakeLagosV2 as the snapshot of a real device, ibm_lagos.

Additionally, the *transpile* tool offered by *Qsikti* was employed, with the optimization_level set to 2, to optimize the circuit topology. For the minimization problem in GEM, the *minimize* package from the *Python* scientific computing library, *scipy* [79], was adapted. Unless stated otherwise, the QHE physical model employs parameters from Table 1. And every simulation result represents the average of five consecutive circuit model simulations, each with 8192 shots.

## 3 Results and discussions

This section introduces three key metrics:

- The population distribution across various energy levels.
- The average power for different cycles.
- The steady-state power and efficiency under varying physical parameters.

The first two metrics are used to assess the feasibility of implementing quantum circuits on quantum computers for the simulation of QHE dynamics and the evaluation of thermodynamic performance. The steady-state performance is applied to examine the regions of physical parameters to ascertain whether this method exhibits expected functionality within these regions.

### 3.1 Populations

The populations obtained from the running quantum circuits and from theoretical calculations are given in Fig. 4. The results from the qasm_simulator (solid points) align well with the theoretical results (line segments). However, the results from FakeLagosV2 (hollow points) deviate slightly due to practical errors. This deviation can be effectively mitigated using GEM (hollow points marked with an X). These findings suggest that this quantum circuit model is capable of theoretical dynamics simulation as well as a promising step towards the the cost-effective application.

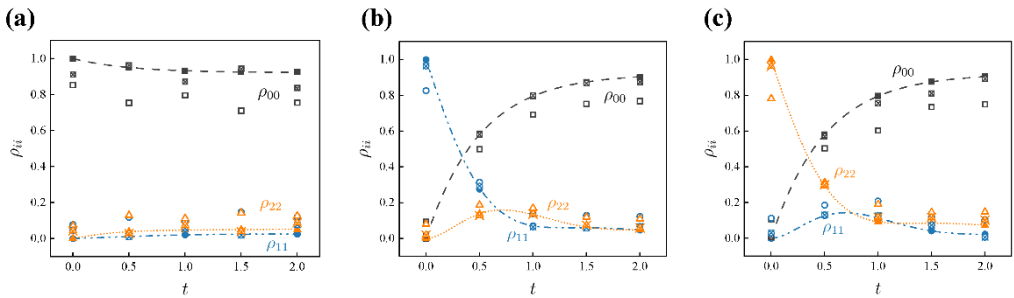

Fig. 4. Temporal change in energy level populations of the three-level QHE. Panels (a), (b), and (c) correspond to different initial states, specifically, $\rho_S(0) = |\varepsilon_0\rangle\langle\varepsilon_0|$, $\rho_S(0) = |\varepsilon_1\rangle\langle\varepsilon_1|$, and $\rho_S(0) = |\varepsilon_2\rangle\langle\varepsilon_2|$, respectively. In each panel, the black dashed

line, blue dash-dotted line, and orange dotted line represent the theoretical populations defined in Eq. (7), i.e., $\rho_{00}$, $\rho_{11}$, and $\rho_{22}$, respectively. The black square, blue circle, and orange triangle symbols represent the corresponding populations obtained from the quantum circuit. Solid symbols denote results from the ideal simulator (qasm_simulator), while hollow symbols represent results from the snapshots of real device (FakeLagosV2). Hollow symbols marked with an 'X' indicate the FakeLagosV2 results after employing GEM.

It's important to note that due to the complexity of this quantum circuit model, the aforementioned circuit have not been executed on a real quantum computer yet. Even after topology optimization, the quantum circuit comprises approximately 170 gates and has a depth of about 120. There are several challenges associated with this:

- The increase in the number of gates within a quantum circuit can lead to the accumulation of errors caused by gate defects, rendering it unsuitable for execution on current quantum computers.
- The increase in circuit depth extends the running time of the circuit, which can amplifies environmental noise and decoherence effects.
- As the number and depth of circuit qubits increase, the effectiveness of GEM diminishes [76].

## 3.2 Average Power

The results outlined in Section 3.1 confirms the feasibility of simulating the dynamics of the three-level QHE by executing quantum circuits. This part employs the method detailed in Section 2.2.2 to construct quantum circuits for the evaluation of the average power (see also Section 2.3). It's important to note that each of these quantum circuits comprises approximately 1600 gates and has a depth of around 1100. Given the considerations outlined in Section 3.1, the quantum circuit runs solely on the qasm_simulator.

Fig. 5 provides the different cycle modes and a comparative analysis of their corresponding average power. Fig. 5(d) demonstrates a good agreement between the results of quantum circuit simulation and numerical calculation. This suggests that, theoretically, Cycle-RL could achieve similar average power when executed on real quantum devices, potentially outperforming other cycles. Although this result is derived form an ideal simulator, it highlights the promising potential of running Cycle-RL on real quantum devices in future studies, with the prospect of achieving a higher average power.

After the evolution, the cycle efficiency can be estimated using the following formula [80]:

$$\eta = \frac{\eta_C}{1 + \frac{\langle \sigma \rangle}{\beta_c \langle P_{he} \rangle}} \tag{39}$$

where, Carnot efficiency $\eta_C = 1 - \beta_h/\beta_c$, $\langle P_{he} \rangle$ is given by Eq. (30), and average entropy production,

$$\langle \sigma \rangle = \tilde{\gamma} \int_0^\infty e^{-\tilde{\gamma} t} \sigma(t) \, dt \tag{40}$$

Here, the instantaneous entropy production is given by:

$$\sigma(t) = \sum_{\alpha = c, h} \beta_\alpha J_\alpha(t) \tag{41}$$

with $J_\alpha(t)$ as instantaneous heat flux. As per Eq. (39), the efficiency derived from numerical calculations stands at 65.34%, while the quantum circuits simulation exhibits an efficiency of 65.39%. This close alignment underscores the precision of the quantum circuit simulation methodologies.

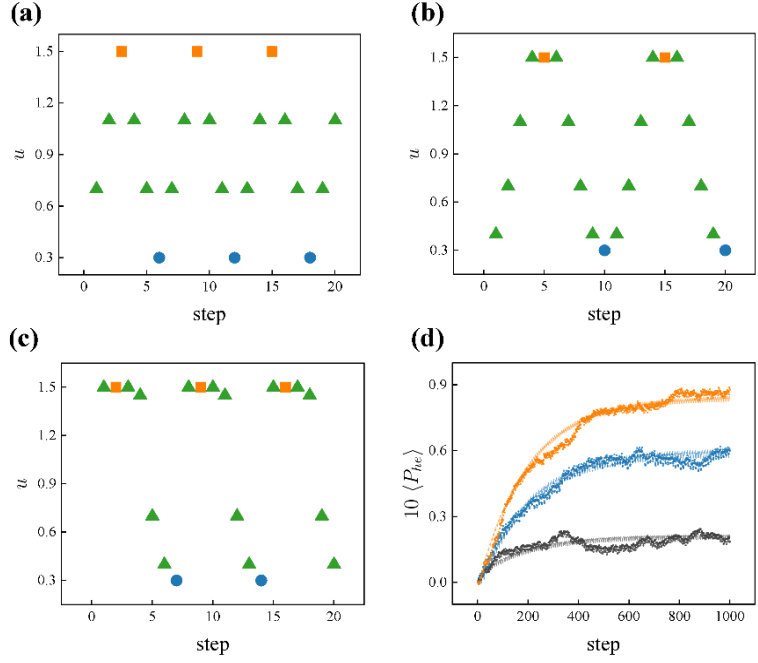

Fig. 5. Different cycle modes and their corresponding average power. (a), (b), and (c) represent different cycle modes, specifically, Cycle-1, Cycle-2, and Cycle-RL. The x-axis denotes the number of steps in the cycle, with each step corresponding to a time length of 0.5. The y-axis represents the system state of the QHE. The orange square, blue circle, and green triangle symbols correspond to the hot, cold, and work processes in the cycle, respectively. (d) Average power of each cycle over 1000 steps. The black, blue, and orange translucent lines (solid dots) represent the results of numerical calculations (quantum circuit simulations) for the average power of Cycle-1, Cycle-2, and Cycle-RL, respectively.

### 3.3 Steady-state performance

This section systematically investigates the stability performance of quantum circuit simulations under a variety of physical parameter settings, utilizing the framework established in Section 2.5. An orthogonal test is employed to perform a parameter sensitivity analysis on steady-state power and efficiency and for the sake of simplicity, the superscript "SS", denoting "steady-state", is omitted in this section.

To mitigate fluctuations caused by potential extreme situations, the number of shots in the quantum circuit simulation is increased from 8192 to 40960. Simultaneously, the number of repetitions for each simulation is maintained at five to ensure experimental consistency. In order to approximate the steady state as closely as possible at the conclusion of the simulation, the evolution time for each case is set according to the principle of $\gamma_h(\varepsilon_{20})t = 4$. Specifically, the evolution time for cases 1, 4, and 7 is eight units, for cases 2, 5, 8 it is four units, and for cases 3, 6, and 9 it is two units. Additionally, data are rigorously screened according to Eq. (38) to eliminate simulation results that fail to meet the normal operating conditions of the heat engine.

Fig. 6 presents both the theoretical steady-state power and the steady-state power obtained through quantum circuit simulation under nine different parameter configurations (cases) as listed in Table 3. It is observed that simulation results (Sim.) align well with the theoretical results (Theo.), even though they exhibit slightly more blank or blue areas. Further comparison of these two results reveals that the blank or blue areas in simulation results generally correspond to the parameter interval associated with lower power. This suggests that the quantum circuit simulation method is largely effective, with the exception of a few scenarios involving low power.

To substantiate this perspective, Fig. 7 provides the relative error between theoretical results and the simulation results. The relative error is defined as follows:

$$Er_P = \frac{\left|(P_{\text{Theo}} - P_{\text{Sim}})/P_{\text{Theo}}\right|}{\left|(P_{\text{Theo}} - P_{\text{Sim}})/P_{\text{Theo}}\right|_{\max}} \tag{42}$$

Here, $P_{\text{Theo}}$ represents the theoretical steady-state power, and $P_{\text{Sim}}$ is the steady-state power obtained through quantum circuit simulation. The subscript "max" means the maximal value. For certain invalid situations like,

$$\begin{aligned} P_{\text{Theo}} > 0, \ P_{\text{Sim}} < 0 \\ P_{\text{Theo}} < 0, \ P_{\text{Sim}} > 0 \end{aligned} \tag{43}$$

The relative error is set as a value slightly greater than 1. It can be observed from Fig. 7 that in the majority of situations, the error is small, which suggests that this method is effective for most scenarios.

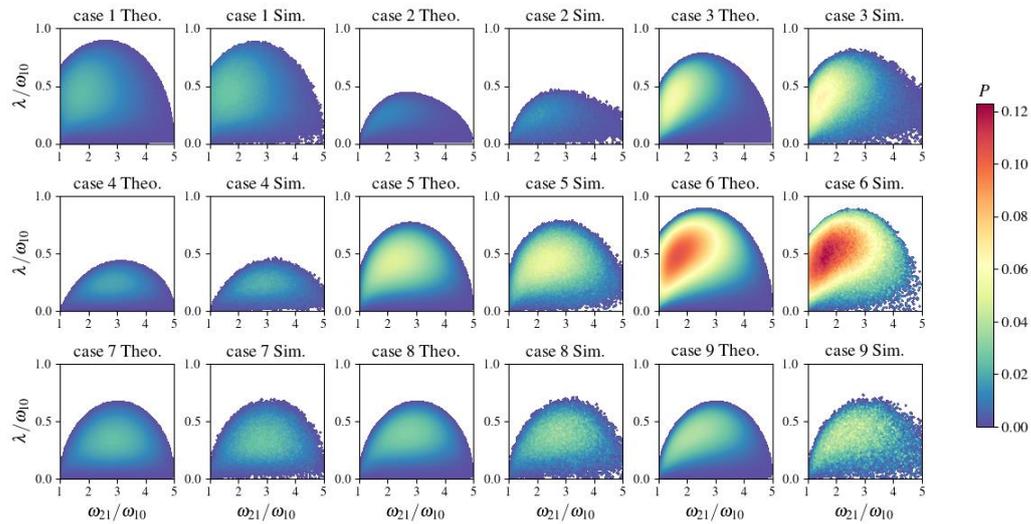

Fig. 6. Theoretical (Theo.) steady-state power and steady-state power obtained through quantum circuit simulations (Sim.) for each case in Table 3.

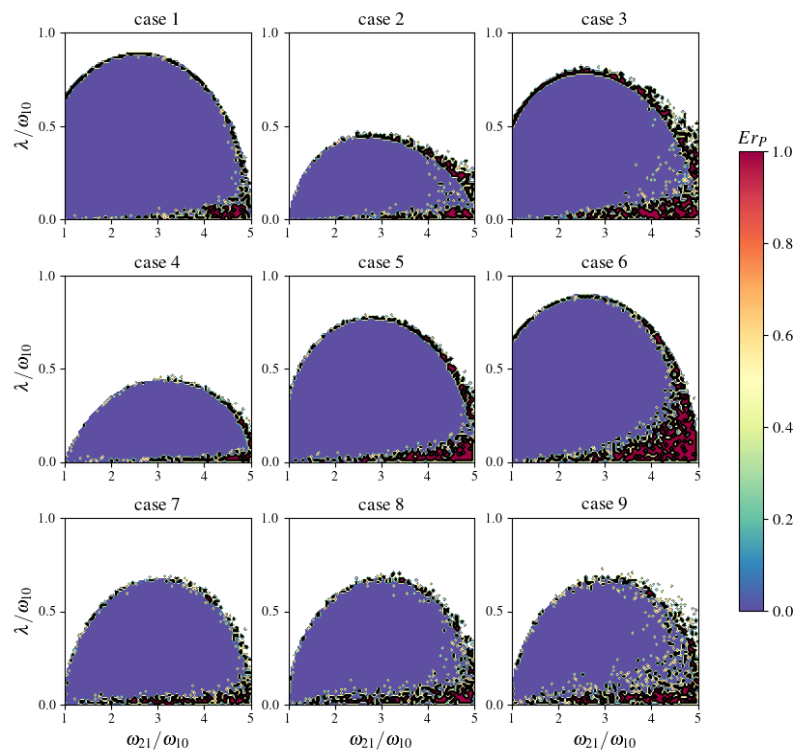

Fig. 7. Relative error between theoretical and simulated steady-state power for each case in Table 3. The black curves denote the invalid situations as defined by Eq. (43).

Similar conclusions can be drawn by comparing the theoretical steady-state efficiency with the simulated steady-state efficiency, as shown in Fig. 8. The solid black line in simulation results (Sim.) demarcates the region that exceeds the Carnot efficiency $\eta_C = 0.8$. These scenarios often align with the lower power regions in Fig. 6 and blank areas in Fig. 8 corresponds to the region where power vanishes. This phenomenon confirms the previous conclusion from another perspective: the prediction accuracy of the quantum circuit simulation method is generally valid except dealing with smaller power situations.

Fig. 9 gives the relative error of steady-state efficiency, defined as:

$$Er_\eta = \frac{\left|(\eta_{\text{Theo}} - \eta_{\text{Sim}})/\eta_{\text{Theo}}\right|}{\left|(\eta_{\text{Theo}} - \eta_{\text{Sim}})/\eta_{\text{Theo}}\right|_{\max}} \tag{44}$$

Invalid situations are defined as:

$$\begin{aligned} \eta_{\text{Theo}} &> 0, \ \eta_{\text{Sim}} < 0 \\ \eta_{\text{Theo}} &< 0, \ \eta_{\text{Sim}} > 0 \\ \eta_{\text{Sim}} &> 0.8 \end{aligned} \tag{45}$$

In these cases, the corresponding relative error is also set as a value slightly greater than 1. A similar conclusion can be drawn from Fig. 9, as black curves appear in invalid power situations.

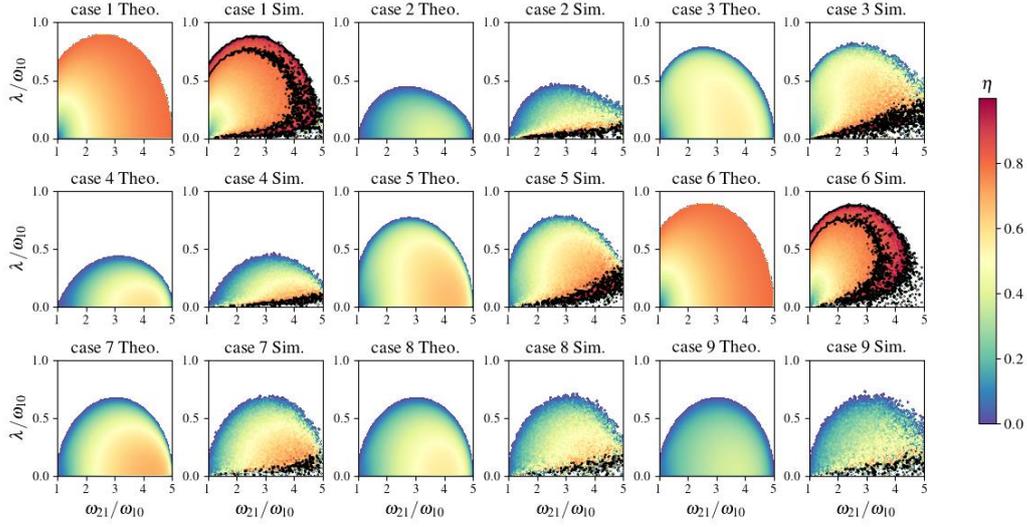

Fig. 8. Theoretical (Theo.) steady-state efficiency and steady-state efficiency obtained through quantum circuit simulations (Sim.) for each case listed in Table 3.

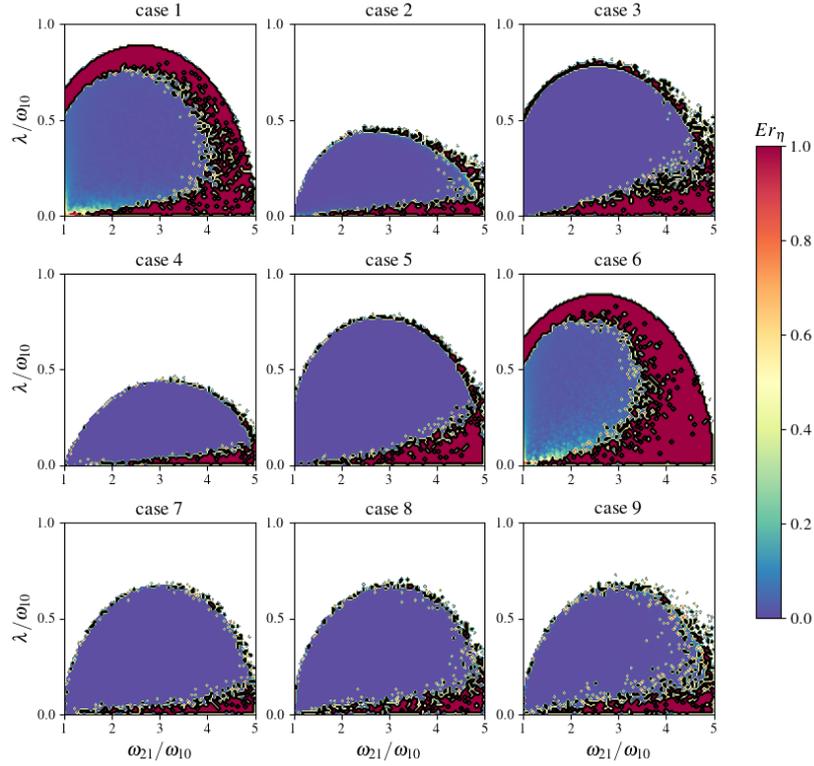

Fig. 9. Relative error between theoretical and simulated steady-state efficiency for each case in Table 3. The black curves denote the invalid situations given by Eq. (45).

## 4 Conclusions

This study applied a novel method for simulating a three-level quantum heat engine (QHE) on quantum circuits. This approach, grounded in the Kraus representation and Sz.-Nagy dilation theorem, offers a promising avenue for reducing experimental costs and gaining valuable insights into practical applications. The initial validation of the dynamics model, which involved monitoring population evolution on both the ideal simulator and snapshots accounting for potential errors, confirmed its feasibility. Furthermore, the average powers evaluation of the QHE across different cycle modes using the thermodynamic model yielded results consistent with theoretical predictions. These outcomes suggest that the benefits, particularly the enhanced average power of the optimized cycle achieved through reinforcement learning, hold substantial potential for real-world application. Finally, a comparison of the steady-state performance between theoretical and quantum circuit simulations suggests that quantum circuit simulations may not be suitable for low-power situations. This result underscores the need for further research and optimization in this area.

**Acknowledgement**

This work was supported by the Taishan Scholar Project (Grand No. tsqn202103142), Natural Science Foundation of Shandong Province (No. ZR2021QE033).